\documentstyle[aps,twocolumn,prb,epsfig]{revtex}

\begin{document}

\title{Enhanced \boldmath ${d_{x^2-y^2}}$ pairing correlations \\
in the two-leg Hubbard ladder}

\author{R.M.\ Noack$^a$, N.\ Bulut$^b$, D.J.\ Scalapino$^b$
and M.G.\ Zacher$^a$}

\address{
a) Institut f\"ur Theoretische Physik,
Universit\"at W\"urzburg, Am Hubland \\
97074 W\"urzburg, Germany}

\address{
b) Department of Physics, University of California \\
Santa Barbara, CA 93106--9530}

\maketitle

\begin{abstract}
The two-leg Hubbard ladder is characterized by the ratio of
the inter- to intra-leg hopping $t_\perp/t$, the relative interaction
strength $U/t$ and the electron filling.
Here, using density matrix renormalization group and Monte Carlo
simulations, we examine the dependence of the pairing correlations on
these parameters.
We find that the pairing correlations
are enhanced when the top of the bonding quasiparticle band and the bottom 
of the antibonding band are near the Fermi level.
We present results on
the single-particle spectral weight and
the antiferromagnetic correlations in order to explain this behavior.
\end{abstract}


One goal of numerical many-body calculations has been to explore the
qualitative properties of a given model, in particular, the nature of the
dominant low-temperature and ground state correlations.
Clearly, a second goal is to determine how to optimize particular
correlations in order to gain further insight into the
mechanism responsible for the correlations,
as well as to provide information which
could be useful in the search for new materials with these
correlations.
Recently, the two-leg Hubbard ladder has been shown to exhibit power law
$d_{x^2-y^2}$-like pair-field 
correlations \cite{Rice,Noack,Balents}.
Here we present a more detailed study of the 
dependence of the these pairing 
correlations upon the basic parameters of the model.
We also examine their relationship to the low-energy single-particle 
spectral weight and the local antiferromagnetic correlations.

The Hamiltonian for the two-leg Hubbard model with near-neighbor hopping
is
\begin{eqnarray}
H & = -t \sum_{i,\lambda\sigma}
\left( c^\dagger_{i\lambda\sigma} c^{}_{i+1\lambda\sigma} + {\rm h.c.}
\right) 
- t_\perp \sum_{i,\sigma} 
( c^\dagger_{i1\sigma}c^{}_{i2\sigma} \nonumber \\
& + {\rm h.c.}) 
+ U \sum_{i\lambda} n_{i\lambda\uparrow} n_{i\lambda\downarrow}.
\label{eq:H}
\end{eqnarray}
Here $c^\dagger_{i\lambda\sigma}$ creates an electron with spin
$\sigma$ on the $i^{\rm th}$ rung of the $\lambda^{\rm th}$ leg,
with $i=1,\ldots, L$ and $\lambda=1$ or 2.
The intra-leg one-electron near-neighbor hopping matrix element is $t$ and 
the inter-leg hopping is $t_\perp$.
The on-site Coulomb interaction is $U$.
We will measure energies in units of $t$ so that the basic parameters of
the two-leg system become the hopping anisotropy $t_\perp/t$, the ratio
$U/t$ of the interaction strength to the intra-leg hopping and the
average electron filling 
$\langle n\rangle=1/(2L) \sum_{i,\lambda} 
\langle n_{i\lambda\uparrow}+n_{i\lambda\downarrow}\rangle$.

\begin{figure}[b]
\vspace*{-1.5cm}
\begin{center}
\epsfig{file=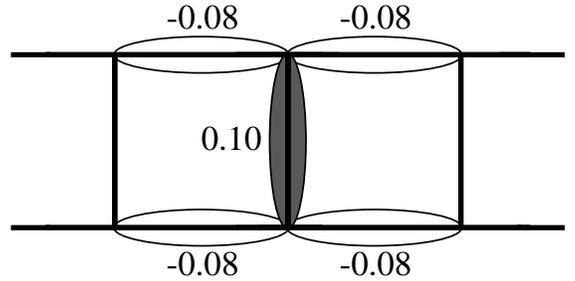, width=6.2cm}
\end{center}
\vspace*{-0.8cm}
\caption{Schematic drawing of the pair-wave function showing the 
values of the off-diagonal matrix element
$\langle N_2| ( c^{\dagger}_{{{\bf r}}\uparrow}
c^{\dagger}_{{{\bf r}}'\downarrow}
-c^{\dagger}_{{{\bf r}}\downarrow}
c^{\dagger}_{{{\bf r}}'\uparrow}) | N_1\rangle$
for creating a singlet pair between near-neighbor sites.
}
\label{fig:amp}
\end{figure}

At half-filling, with $U/t>0$, the two-leg Hubbard ladder is found to
have both a charge gap and a spin gap \cite{Noack}. 
Thus it is insulating with exponentially decaying short-range
antiferromagnetic correlations.
When the ladder is doped away from half-filling,
the charge gap disappears, but a reduced spin gap remains for a range of
doping and $t_\perp/t$ values.
In addition, $d_{x^2-y^2}$-like 
power-law correlations are observed \cite{Rice,Noack,Balents}.
The internal structure of the pairs is 
illustrated in Fig. \ref{fig:amp}, in which we
show the amplitude
\begin{equation}
\left\langle N_2 \left| \left(
c^\dagger_{{\bf r}\uparrow} c^\dagger_{{\bf r}'\downarrow} -
c^\dagger_{{\bf r}\downarrow}c^\dagger_{{\bf r}'\uparrow} 
\right) \right| N_1 \right\rangle 
\label{eq:amp}
\end{equation}
for adding a singlet pair on near-neighbor sites along and across the
legs, calculated using 
the density matrix renormalization group (DMRG) method.
Note the $d_{x^2-y^2}$-like change in sign of this matrix element.
Here $|N_1\rangle$ is the ground state with four holes relative to the
half-filled band and $|N_2\rangle$ is the ground state with two holes on
a $2\times16$ ladder.
For $U/t=8$ and $t_\perp/t=1.5$, 
the $d_{x^2-y^2}$-like structure of
the matrix element, 
Eq.~(\ref{eq:amp}), extends
over a region $|{\bf r}-{\bf r}'|$ of order 4 rungs.

\begin{figure}
\begin{center}
\epsfig{file=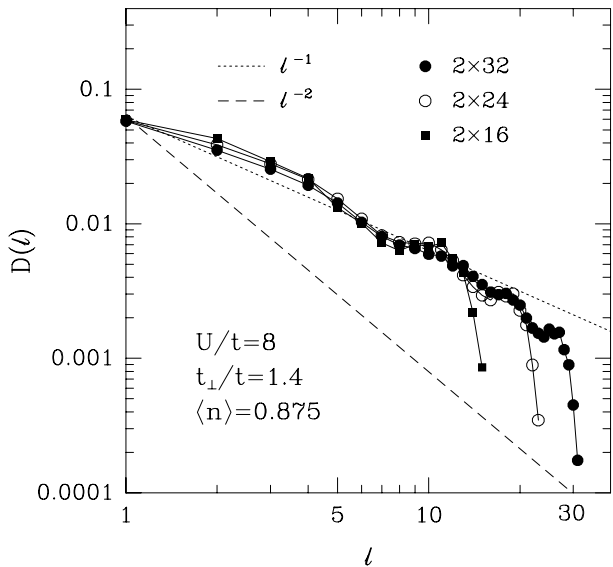, height=8.5cm}
\end{center}
\caption{The rung-rung pair-field correlation function $D(\ell)$ 
versus $\ell$ for $U/t=8$, $t_{\perp}/t=1.4$,
$\langle n\rangle=0.875$ on $2\times L$ ladders with 
$L=16$, 24 and 32.
The dashed and the dotted lines vary as $\ell^{-2}$ and $\ell^{-1}$,
respectively.
}
\label{fig:Dell}
\end{figure}

DMRG techniques \cite{White1}
have also been used to calculate the ground state expectation 
value of the rung-rung pair-field correlation function
\begin{equation}
D(i,j) = \langle \Delta(i)\Delta^\dagger(j) \rangle. 
\label{eq:Dell}
\end{equation}
Here
\begin{equation}
\Delta^\dagger(i) = 
\left(
c^\dagger_{i1\uparrow}c^\dagger_{i2\downarrow} -
c^\dagger_{i1\downarrow}c^\dagger_{i2\uparrow} \right) 
\label{eq:delta}
\end{equation}
creates a singlet pair across the $i^{\rm th}$ rung.
For technical reasons, the DMRG calculations are carried out for
$2\times L$ ladders with open boundary conditions.
For this reason, to obtain information on the spatial decay of the
pairing correlations, we have averaged the pair-field correlation
function $D(i,j)$ over (typically) six $(i,j)$ pairs with $\ell = |i-j|$
fixed.
This averaging procedure starts with symmetrically placed $(i,j)$ values
and then proceeds to shift these to the left and right of center.
When $|i-j|$ approaches the lattice size, the number of possible
$(i,j)$ pairs
is reduced due to the proximity to the boundaries.
Results for $D(\ell=|i-j|)$ computed in this way for 
$2\times 16$, $2\times 24$ and $2\times 32$
lattices with $t_\perp/t = 1.4$, $U/t = 8$, and 
$\langle n\rangle =0.875$ are shown
in Fig. \ref{fig:Dell}.
The dotted line corresponds to a power law decay $\ell^{-1}$.
As can be
seen, there are clear finite size effects when $\ell$ approaches $L$.
However, one can determine when this occurs by comparing ladders of
increasing length.
In the following, we will discuss results obtained for $2\times32$
ladders over distances $\ell\leq 20$, for which the end effects are
negligible.

\begin{figure}
\begin{center}
\epsfig{file=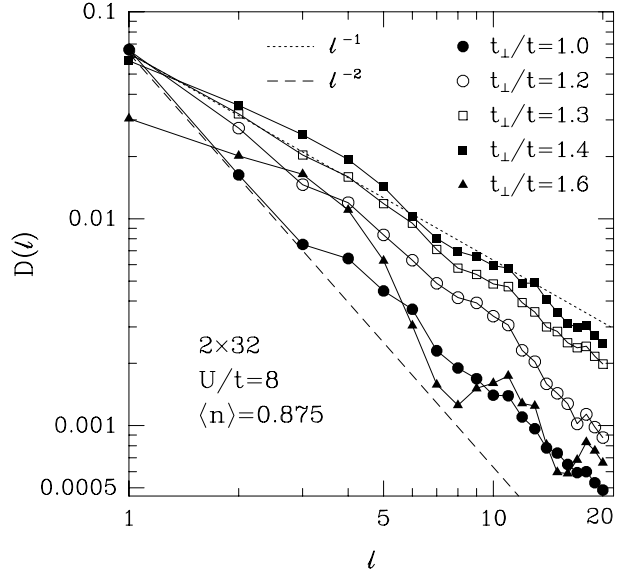, height=8.5cm}
\end{center}
\caption{$D(\ell)$ versus $\ell$ for various values of
$t_{\perp}/t$ with $U/t=8$ and $\langle n\rangle=0.875$.
}
\label{fig:Delltp}
\end{figure}

\begin{figure}[b]
\vspace*{-0.8cm}
\begin{center}
\epsfig{file=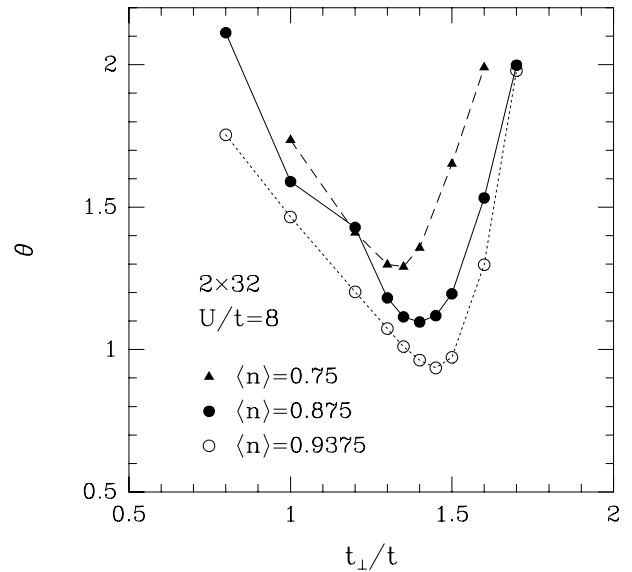, height=8.5cm}
\end{center}
\caption{Exponent $\theta$ versus $t_{\perp}/t$ for 
$U/t=8$ at fillings $\langle n\rangle = 0.75$, 0.875 and 0.9375.
}
\label{fig:exponent}
\end{figure}

In Fig. \ref{fig:Delltp}
we show $D(\ell)$ versus $\ell$ for various values of
$t_\perp/t$ with $U/t=8$ and $\langle n\rangle=0.875$.
As this log-log plot shows, the pair-field correlations exhibit
behavior consistent with a
power-law decay
\begin{equation}
D(\ell) \sim {1\over\ell^\theta}
\label{eq:Dell2}
\end{equation}
\noindent for $t_\perp/t < 1.6$.
For $t_\perp/t = 1.6$, $D(\ell)$ shows large oscillations and markedly
reduced strength which we believe is associated with a transition to
a phase in which the antibonding band is unoccupied.
In Fig. \ref{fig:exponent} we show 
the exponent $\theta$ versus $t_\perp/t$
obtained from a linear least-squares 
\noindent fit 
of $\log D$ versus $\log(\ell)$
for $\ell=1$ to 18 at various fillings.
Because of the presence of some oscillations in $D(\ell)$ and the
finite range of $\ell$ available, the values of $\theta$ thus obtained
can only be rough estimates, but they do allow 
one to examine 
the trend in $\theta$ as $t_{\perp}/t$ is varied.
We find that $\theta$ initially decreases with increasing 
$t_{\perp}/t$ until $t_{\perp}/t$ becomes of order 1.3  to 1.4
for $U/t=8$ and the fillings shown.
For larger values of $t_{\perp}/t$, 
we find that the 
antibonding band is no longer occupied and the 
pairing 
correlations rapidly collapse.
Thus the pairing correlations are enhanced near the 
point at 
which the antibonding band moves through the Fermi 
level \cite{Noack,Yamaji,Kuroki}. 
  
\begin{figure}
\begin{center}
\epsfig{file=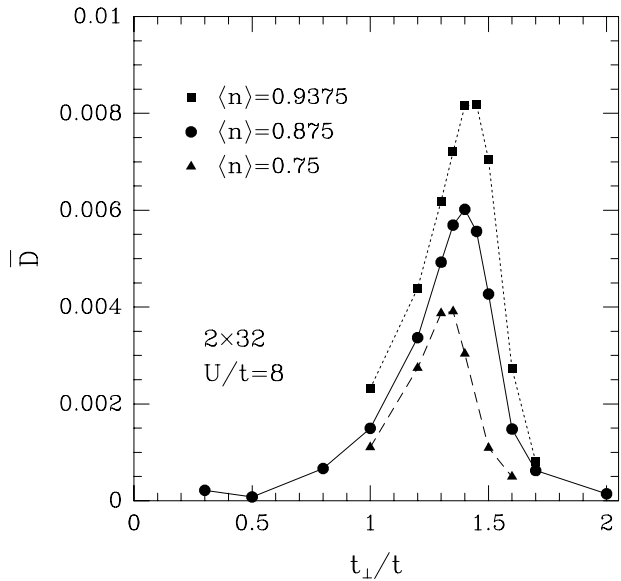, height=8.5cm}
\end{center}
\caption{$\overline{D}$ versus $t_{\perp}/t$ for 
$U/t=8$ at fillings $\langle n\rangle = 0.75$, 0.875 and 0.9375.
}
\label{fig:Dbar}
\end{figure}

\begin{figure}
\begin{center}
\epsfig{file=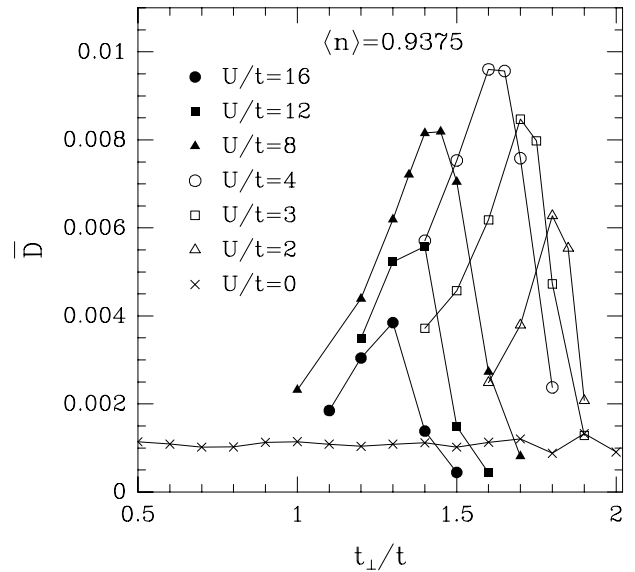, height=8.5cm}
\end{center}
\caption{$\overline{D}$ versus $t_{\perp}/t$ for various values of 
$U/t$ at filling $\langle n\rangle = 0.9375$.
}
\label{fig:Dbartp}
\end{figure}

\begin{figure}
\vspace*{-0.8cm}
\begin{center}
\epsfig{file=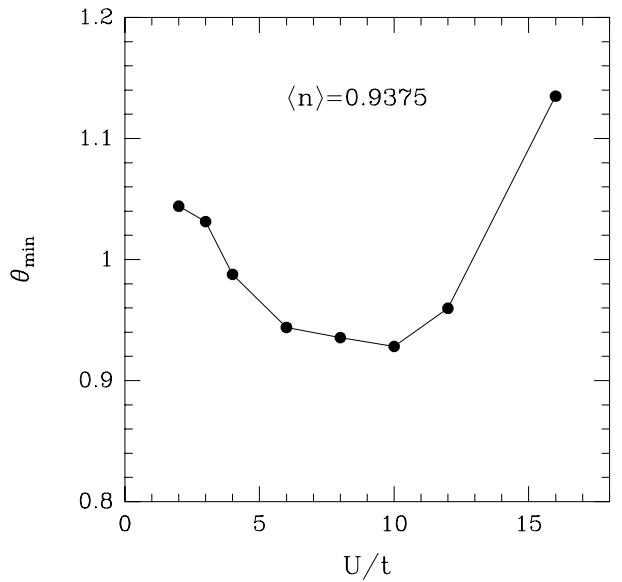, height=8.5cm}
\end{center}
\caption{
Minimum value of the exponent, $\theta_{\rm min}$,
versus $U/t$ at filling $\langle n\rangle=0.9375$.
}
\label{fig:thetamin}
\end{figure}

Another measure of the strength of the pair field correlations,
which we will use, is the average of $D(\ell)/D(1)$ for rung separations 
$\ell = 8$ to 12:
\begin{equation}
\overline D \equiv
{\textstyle{1\over5}} \sum^{12}_{\ell=8} {D(\ell)\over D(1)}.
\label{eq:Dbar}
\end{equation}
At a distance $\ell=8$, we have moved beyond the correlation length
characterizing the size of a pair and are probing the pair
center-of-mass correlations.
Figure \ref{fig:Dbar} shows $\overline D$ versus
$t_\perp/t$ for $U=8\,t$ at fillings 
$\langle n\rangle=0.75$, 0.875, and 0.9375.
Again, the pairing response is seen to initially increase with $t_\perp/t$,
reach a peak value, and then rapidly fall off.
The variation of $\overline D$ with $U/t$ 
is shown in Fig. \ref{fig:Dbartp}.
Here we see that the peak value of $\overline D$ increases with $U/t$
reaching a broad maximum for $U/t$
of order 3 to 8.
In addition, the value of $t_\perp/t$ at which the peak occurs shifts
towards smaller values as $U/t$ increases.
Also shown is $\overline D$ calculated for $U=0$ on the infinite
ladder, showing the strong enhancement of the pairing correlations 
for the interacting system 
over those of the noninteracting system.
In Fig. \ref{fig:thetamin} we show results on how the minimum
value of the exponent, $\theta_{\rm min}$, varies as a function
of $U/t$ at $\langle n\rangle=0.9375$.
Here, $\theta_{\rm min}$ is 
obtained by varying $t_{\perp}/t$ 
while keeping $U/t$ fixed.
  
\begin{figure}
\begin{center}
\epsfig{file=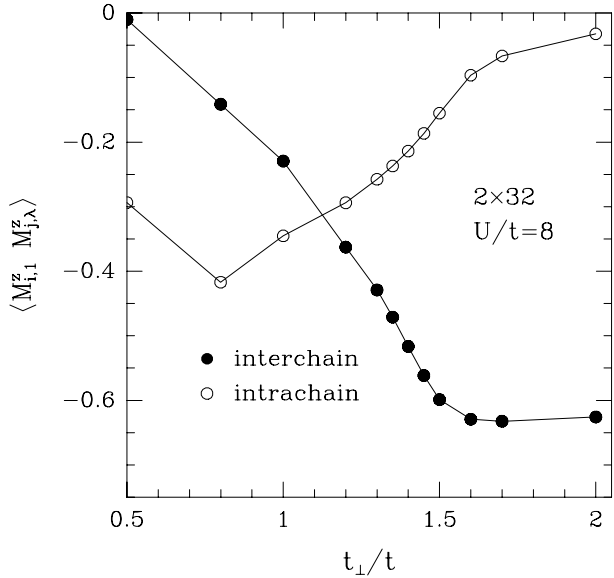, height=8.5cm}
\end{center}
\caption{The inter- ($j=i$, $\lambda=2$) 
and intra-chain ($j=i+1$, $\lambda=1$)
near-neighbor magnetic correlation function 
$\langle M_{i,1}^z M_{j,\lambda}^z\rangle$ versus $t_{\perp}/t$ for 
$\langle n\rangle=0.875$ and $U/t=8$.
}
\label{fig:mag}
\end{figure}

{}From the results shown in 
Figs. \ref{fig:Dbar} through \ref{fig:thetamin}, 
it is clear that the strength of the
pairing correlations as measured 
by $\overline D$,
Eq.~(\ref{eq:Dbar}), 
depend sensitively on $t_\perp/t$ and the filling 
$\langle n\rangle$ as
well as on $U/t$.
The fact that $\overline D$ decreases for large
\noindent 
values of $U/t$ is consistent 
with the variation of the exchange interactions 
$J\cong 4\,t^2/U$ and 
$J_\perp\cong 4\,t^2_\perp/U$, 
which become weaker at large values of $U$.
It is interesting to note that $\overline D$ has its largest value for 
intermediate coupling $U/t$.
Since $J_\perp/J$ varies as $(t_\perp/t)^2$, 
the interchain antiferromagnetic correlations are enhanced relative to
the intrachain correlations as $t_\perp/t$ increases.
The $z$--$z$ spin correlation function
$\langle M^z_{i,1} M^z_{j,\lambda}\rangle$, where
$M^z_{i,\lambda} = n_{i,\lambda \uparrow} - n_{i,\lambda \downarrow}$,
is shown for interchain ($j=i$, $\lambda=2$) and intrachain 
($j=i+1$, $\lambda=1$) nearest neighbor sites in Fig. \ref{fig:mag}.
Thus both $U/t$ and $t_\perp/t$ enter in determining the strength and
anisotropy of the exchange couplings and the local structure of the
antiferromagnetic correlations on the ladder.
The underlying pairing interaction is clearly associated with these
short-range antiferromagnetic correlations.

\begin{figure}[t]
\begin{center}
\epsfig{file=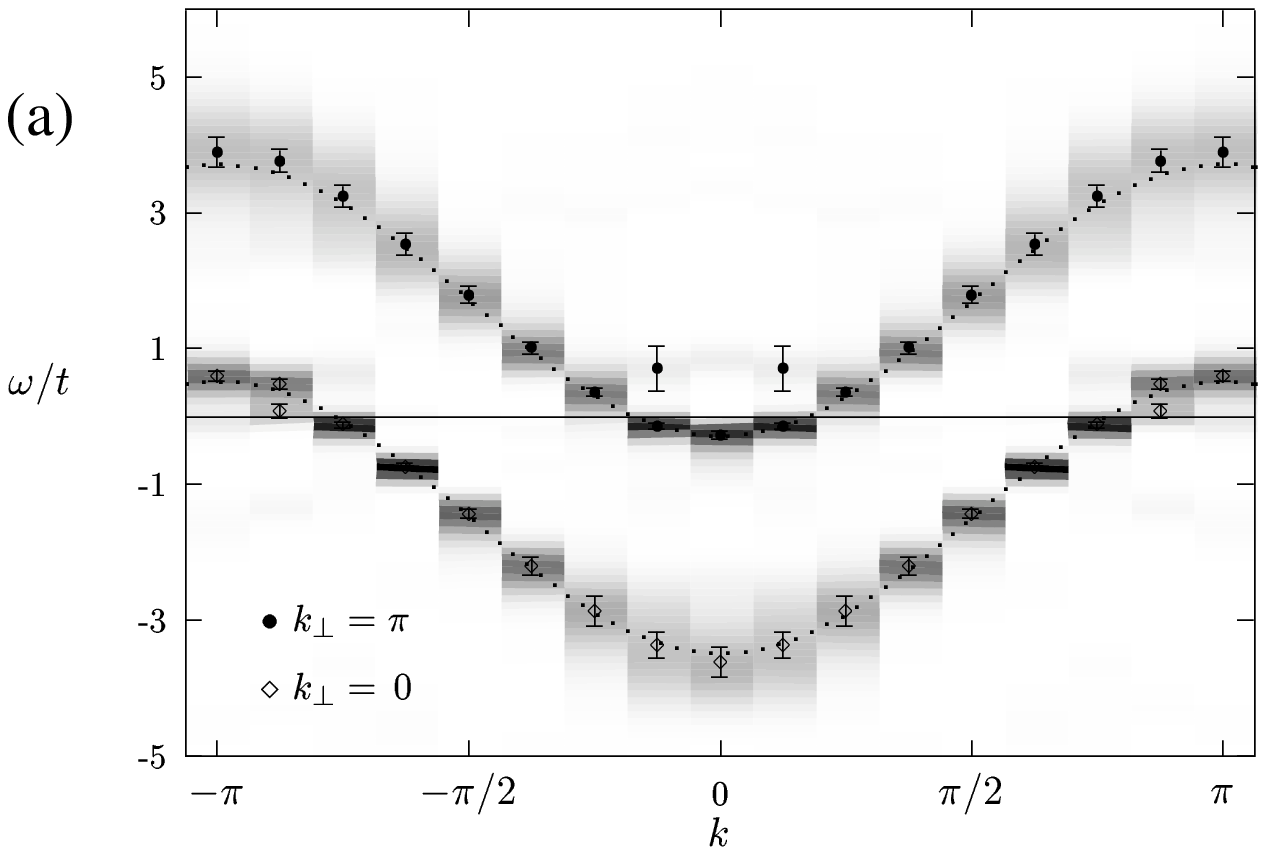, height=5.7cm}
\end{center}
\vspace*{-0.3cm}
\begin{center}
\epsfig{file=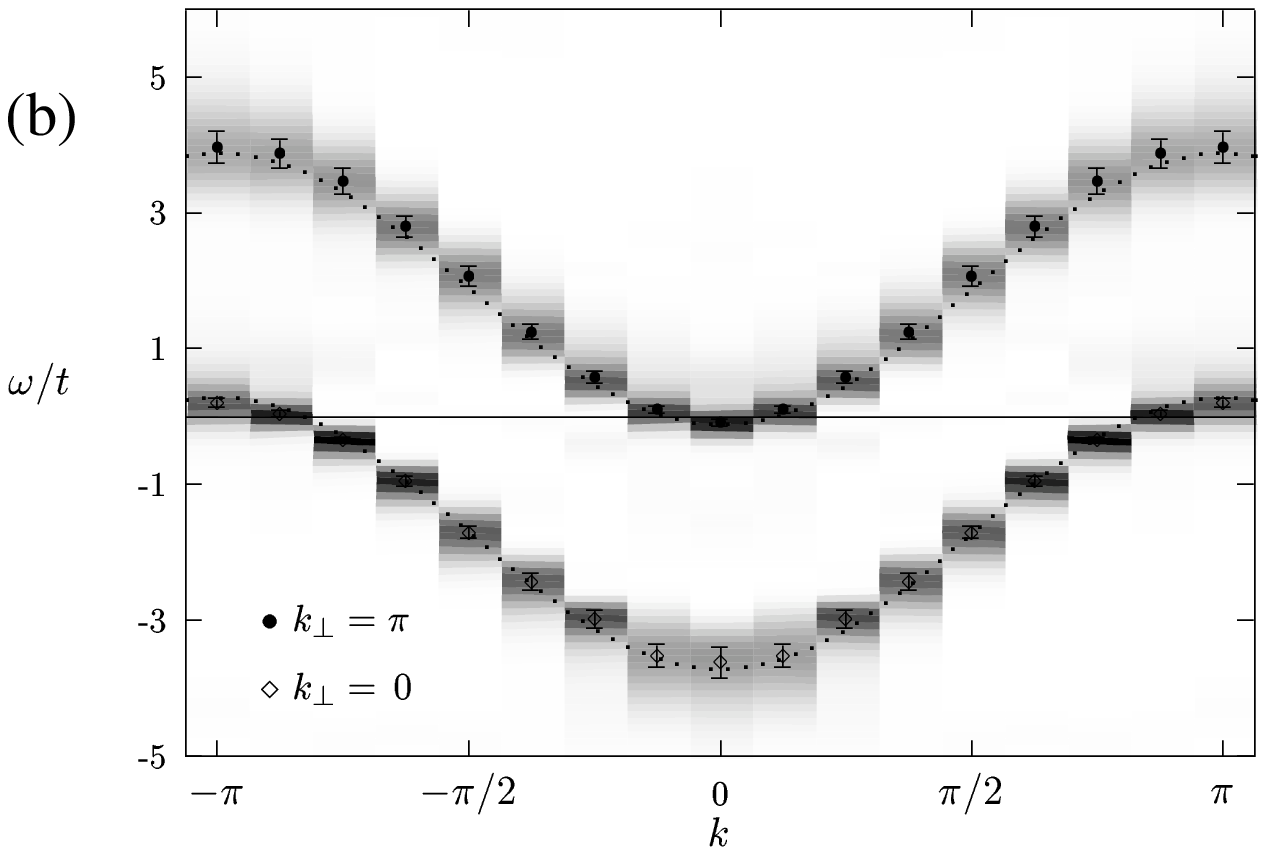, height=5.7cm}
\end{center}
\vspace*{-0.3cm}
\begin{center}
\epsfig{file=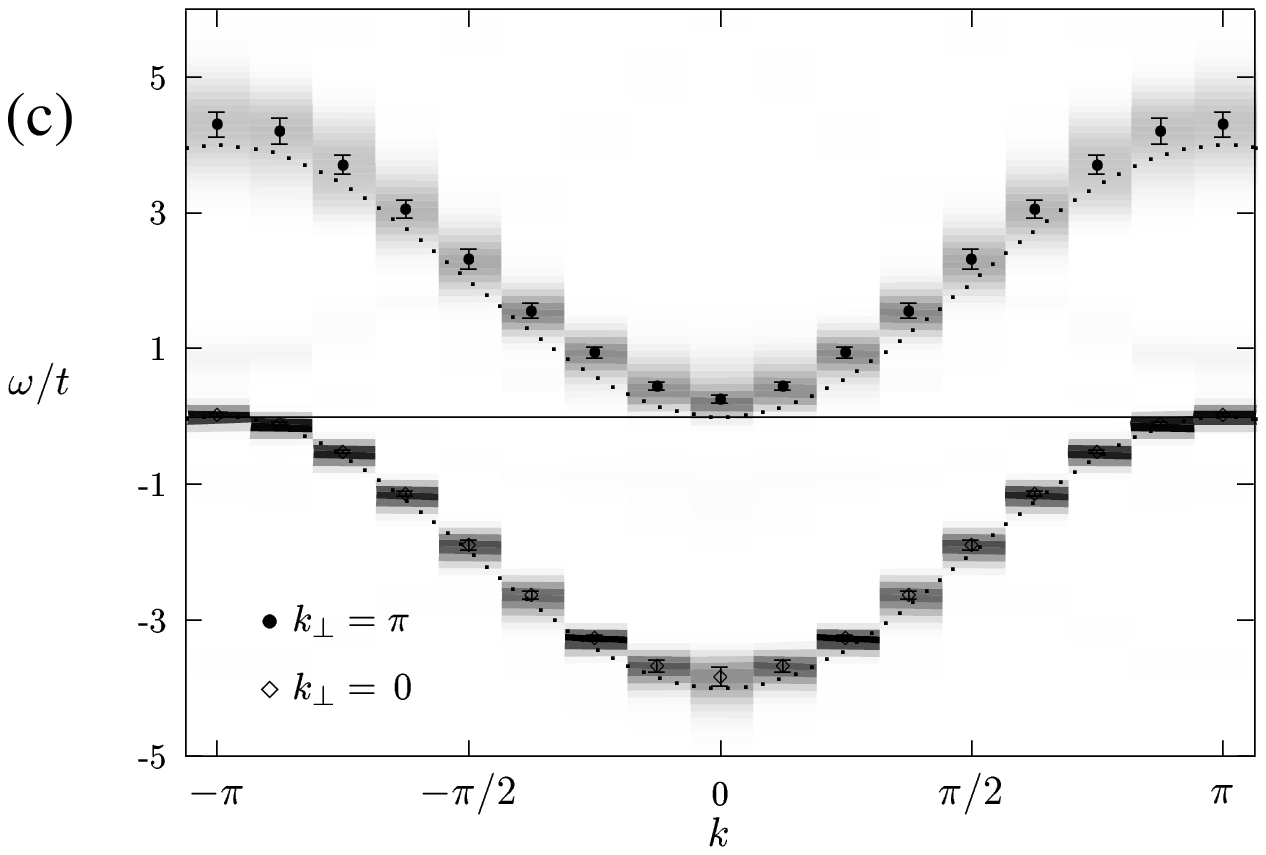, height=5.7cm}
\end{center}
\caption{Single--particle spectral weight $A({{\bf k}},\omega)$ versus 
  $\omega$ at different values of ${{\bf k}}$ for 
  $U/t=2$, $\langle n\rangle=0.94$ 
  and a temperature $T/t=0.125$.
  Here results are shown for (a) $t_{\perp}/t=1.6$,
  (b) 1.8 and (c) 2.0.
  The bonding ($k_\perp=0$) and antibonding ($k_\perp=\pi$) branches are
  superimposed on the density plot, with the indicated symbols marking
  the positions of peaks associated with each band.
  Areas of darker shading
  correspond to areas of higher relative spectral weight.
  The dotted lines indicate the position of the noninteracting ($U=0$)
  bands and the solid line at $\omega=0$ indicates the Fermi level.
}
\label{fig:apwU2}
\end{figure}

In addition, and of particular importance, the ratio $t_\perp/t$, the
filling $\langle n\rangle$, and the interaction strength 
$U/t$ determine the energy
and momentum structure of the single-particle spectral weight
\begin{equation}
A({{\bf k}},\omega) = -{1\over\pi} {\rm Im}\, G({{\bf k}},\omega). 
\label{eq:apw}
\end{equation}
Results for $A({\bf k},\omega)$, obtained 
from a maximum entropy analytic continuation
of Monte Carlo data for a $2\times16$ lattice with periodic boundary
conditions, are shown in Fig. \ref{fig:apwU2} for $U/t=2$ and in 
Fig.\ 10 for $U/t=4$.\cite{Endres}
Here, we have chosen $\langle n\rangle=0.94$, the filling at which
the pairing correlations have maximum strength in Fig.\ \ref{fig:Dbar},
and a temperature $T=0.125t$, which is sufficiently low to ensure
that the band structure would not shift at still lower temperatures.
(Note that $\langle n \rangle$ is a measured quantity in the grand
canonical Monte Carlo simulations, and is thus not fixed as in
the DMRG calculations for which the number of particles is definite.)
We show data for the $k_\perp=0$ (bonding) and $k_\perp=\pi$
(antibonding) branches superimposed in a
density plot in which the density of the shading 
represents the
relative amount of spectral weight and is plotted as a function of the
momentum along the chains, $k$, and the energy $\omega$.
Although the resolution is limited by the finite temperature and 
statistical errors, one can see coherent, dispersive bonding and 
antibonding bands.

The spectral weight distribution clearly evolves with
$t_{\perp}/t$ for both values of $U/t$.
For $U/t=2$, the width and dispersion of the bands follow quite
closely those of the noninteracting, $U=0$, bands, which are
indicated by dotted lines in Fig.\ \ref{fig:apwU2}.
As $t_{\perp}/t$ increases from 1.6 to 2.0, the peaks 
in the bonding and antibonding spectral weight at the Fermi surface 
move towards a momentum separation
$\Delta{\bf k}=(\pi,\pi)$ where scattering from the short range 
antiferromagnetic correlations can become most effective.
The point at which the bottom of the antibonding band is just at the
Fermi level, $t_{\perp}/t=1.8$, (Fig.\ \ref{fig:apwU2}(b)) agrees well
with the position of the peak
in $\overline{D}$ in the $U/t=2$ curve in Fig.\ \ref{fig:Dbartp}.
At $t_{\perp}/t=1.8$, there is large amount of single-particle spectral 
weight near the Fermi level in the bonding band near $k=\pi$ and in
the antibonding band near $k=0$.
At larger values of $t_{\perp}/t$,
the antibonding band pulls away from
the Fermi energy, leaving only the bonding band with spectral weight at
the Fermi energy,
as shown in Fig.\ \ref{fig:apwU2}(c) for $t_{\perp}/t=2.0$.
We believe that it is the variation in spectral weight with $t_\perp$,
coupled to antiferromagnetic fluctuations which are strongly peaked 
near $(\pi,\pi)$ that is primarily responsible for the peak in 
$\overline D$ versus $t_\perp/t$

\begin{figure}[t]
\begin{center}
\epsfig{file=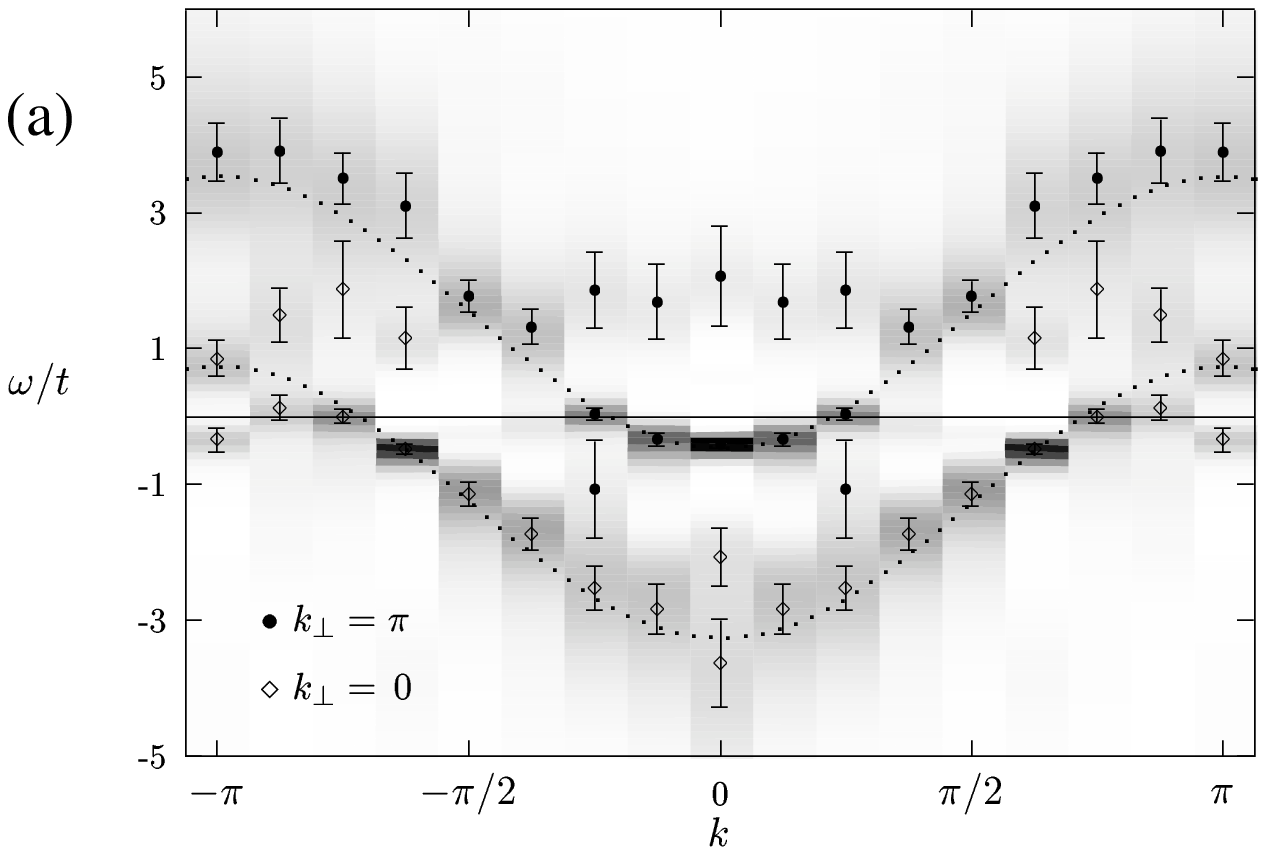, height=5.7cm}
\end{center}
\begin{center}
\epsfig{file=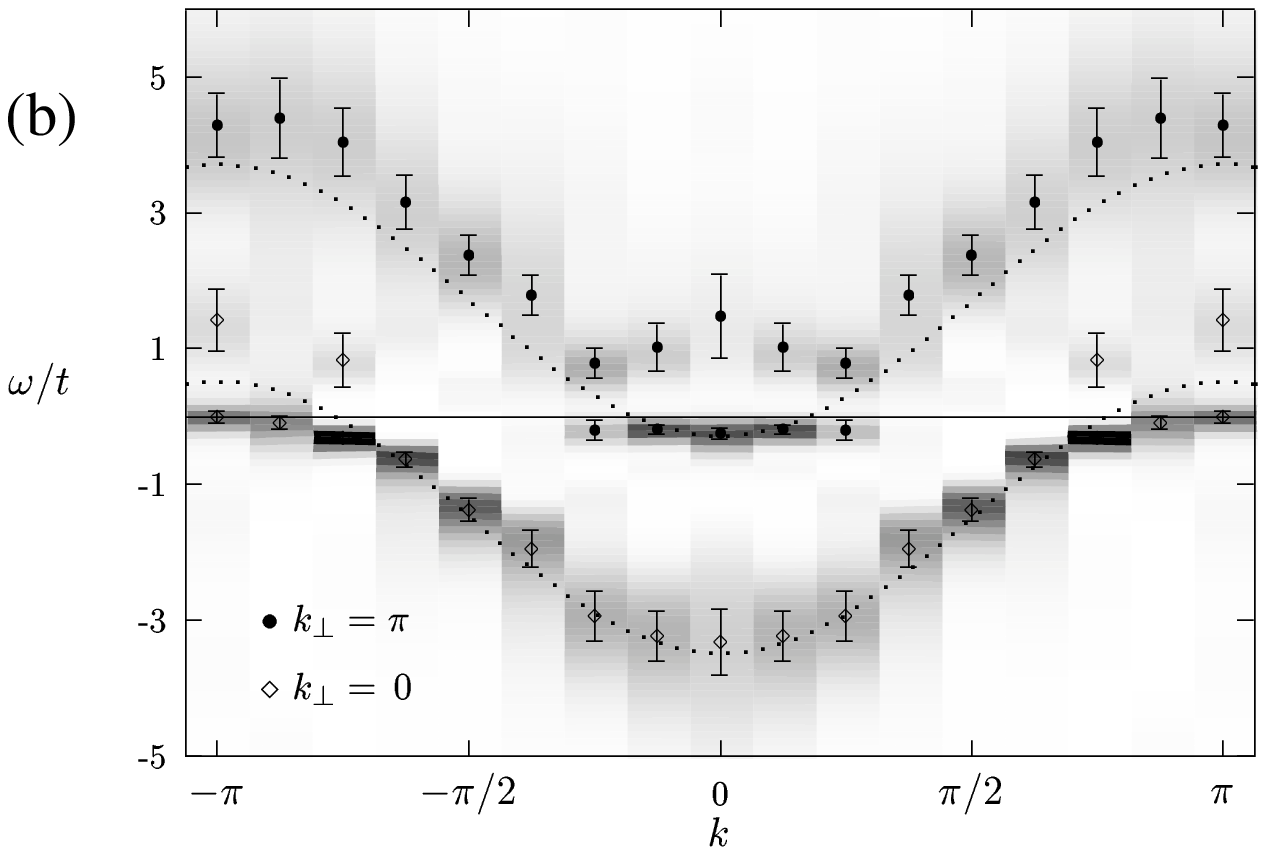, height=5.7cm}
\end{center}
\begin{center}
\epsfig{file=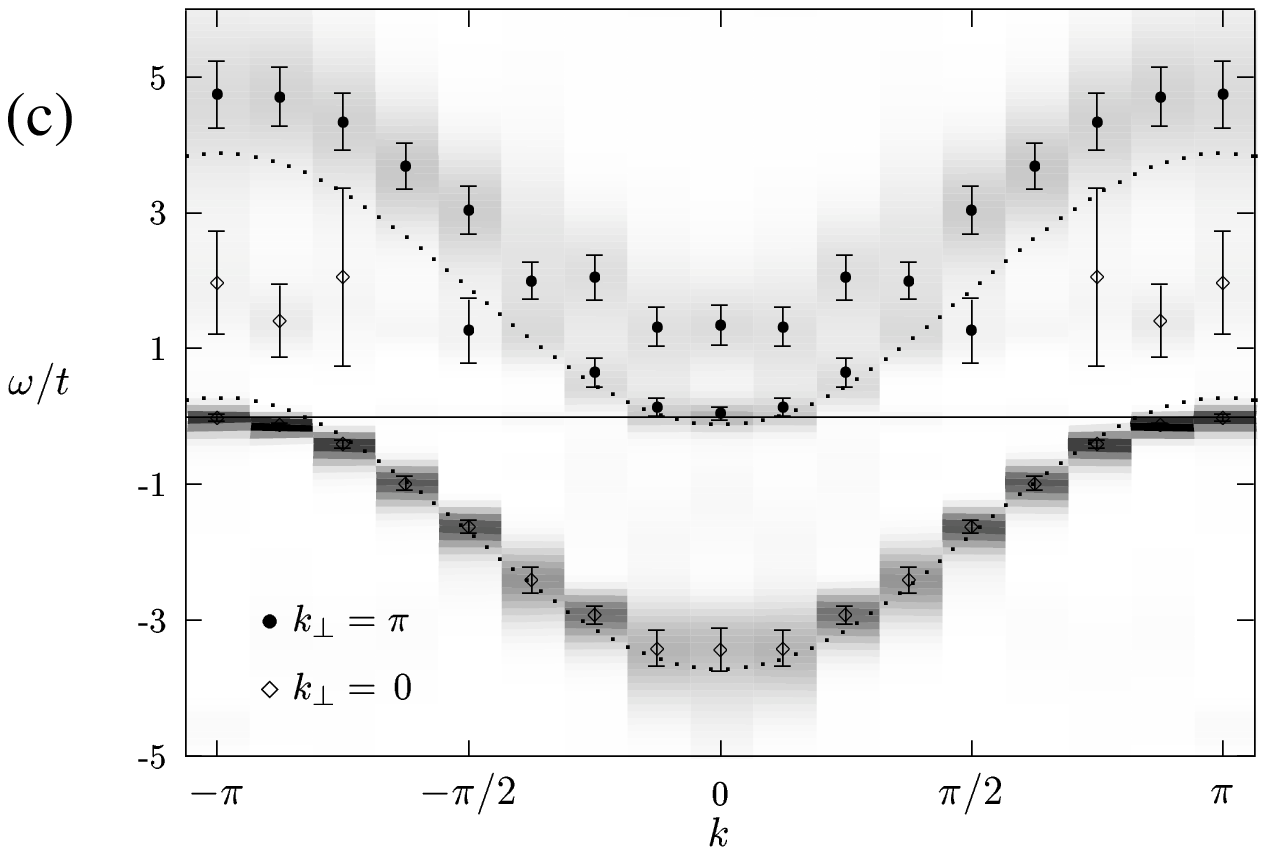, height=5.7cm}
\end{center}
\caption{Single--particle spectral weight $A({{\bf k}},\omega)$ versus
${\bf k}$ and $\omega$ plotted as in Fig.\ 9 for the same parameters 
as in Fig.\ 9, except with $U/t=4$ and (a) $t_{\perp}/t=1.4$, (b) 1.6 
and (c) 1.8.  The dotted lines indicate the position of the 
noninteracting ($U=0$) bands and the solid line at $\omega=0$ indicates 
the Fermi level.
}
\label{fig:apwU4}
\end{figure}

As shown in Fig.\ \ref{fig:Dbartp}, 
as the on-site coupling is increased to $U/t=4$, the value of
$t_\perp/t$ at which $\overline D$ peaks shifts to smaller values.
This is due to narrowing of the quasi-particle bands relative to the
$U=0$ bands for larger $U$.
In Fig.\ \ref{fig:apwU4}, we exhibit the single-particle spectral
weight at $t_\perp/t=1.4$, 1.6, and 1.8, values which bracket the peak in
$\overline D$, which occurs at $t_\perp/t=1.6$.
As in Fig.\ \ref{fig:apwU2}, one can clearly see that the maximal
spectral weight at the 
Fermi level in the antibonding band at $k=0$ and the bonding band
at $k=\pi$ occurs just at the value
of $t_\perp/t$ associated with the peak.
One can also see additional
structures in both bands, which are reflections of the portion of each
band below the Fermi level about the Fermi level.
These structures, present due to stronger antiferromagnetic
correlations at the larger $U/t$ value, are remnants of features that would
be generated by the halving of the
Brillouin zone in an antiferromagnetically ordered state.
Both bands are flattened relative to the $U=0$ bands 
near the Fermi level, i.e. at $k=0$ in the antibonding
band and at $k=\pi$ in the bonding band,
especially for $t_{\perp}/t=1.6$ 
which is 
shown in Fig.\ \ref{fig:apwU4}(b).
As a result of this,
there is large amount of single-particle spectral weight 
near the fermi level.
This behavior is similar to the buildup of spectral weight near
$(\pi,0)$ observed in simulations of the two-dimensional Hubbard model
doped near half-filling \cite{BSW,Preuss}.

These results show that near half-filling, the strength
of the pairing correlations depends
sensitively upon $t_\perp/t$ and the band filling as well as $U/t$.
The optimum value of $\overline D$ occurs for intermediate 
values of $U/t$ and for $\langle n\rangle$ near half-filling
with $t_\perp/t\simeq 1.5$.
We find that for these values, there are strong inter-chain
antiferromagnetic correlations and a large single-particle spectral
weight at the Fermi surface points of the bonding and antibonding bands.

\acknowledgments

This research was supported in part by the National Science Foundation
under Grant No. DMR95-27304 and PHY94-07194 (N.B. and D.J.S.).
M.G.Z. would like to acknowledge support from the Bavarian FORSUPRA II
program.
Some of the numerical calculations reported here were carried out at
the Cray--T90's at the LRZ in M\"unchen and the HLRZ in J\"ulich.
N.B. and D.J.S. gratefully acknowledge computational support
from the San Diego Supercomputer Center.

\end{document}